\documentclass[%
 reprint,
superscriptaddress,
 amsmath,amssymb,
 aps,
]{revtex4-2}

\usepackage{graphicx}
\usepackage{dcolumn}
\usepackage{bm}
\usepackage[caption=false]{subfig}
\newcommand*{\Scale}[2][4]{\scalebox{#1}{$#2$}}%
\begin{document}
\preprint{APS/123-QED}
\title{The free energy balance equation applied to gyrokinetic instabilities, the effect of the charge flux constraint, and application to simplified kinetic models}

\author{M. Kotschenreuther}
\email[Email: ]{mtk@austin.utexas.edu}
\affiliation{University of Texas, Austin, Texas, USA}

\author{X. Liu}%
\affiliation{University of Texas, Austin, Texas, USA}%

\author{S. M. Mahajan}%
\affiliation{University of Texas, Austin, Texas, USA}%

\author{D. R. Hatch}%
\affiliation{University of Texas, Austin, Texas, USA}%

\begin{abstract}
The free energy balance equation for gyrokinetic fluctuations is derived and applied to instabilities. An additional term due to electromagnetic sources is included. This can provide a simpler way to compute the free energy balance in practical applications, and is also conceptually clarifying. The free energy balance, by itself, is not sufficient to determine an eigenfrequency. The preceding results are derived in general geometry. The charge flux constraint in gyrokinetics can provide a necessary additional relation, and the combination of these two can be equivalent to a dispersion relation. The charge flux constraint can prevent the appearance of an unstable eigenmode even though the free energy balance would allow strongly growing fluctuations. The application of these concepts to simplified kinetic models in simplified geometry is also indicated. 

\end{abstract}

\maketitle

\section{Derivation of the gyrokinetic free energy balance equation, and its application to instabilities}

Instabilities that are described by the gyrokinetic equation are typically the main source of transport in magnetically confined plasmas. The gyrokinetic equation describes small fluctuation around a local Maxwellian that has relatively small gradients of temperature and density (in comparison to the fluctuation scales) \cite{hazmei}. The  preceding sentence describes an ideal situation in which to apply the concepts of non-equilibrium thermodynamics, where it is often very advantageous that the departures from a Maxwellian are small. Surprisingly, such concepts have not been widely used to understand gyrokinetic instabilities and transport. But we will see that the use of fundamental non-equilibrium thermodynamic relationships is extremely illuminating. Here, we derive and describe a free energy balance equation that describes the conversion of free energy in equilibrium gradients into the free energy of the fluctuations. Unlike other derivations, we include potential source terms in the equation. This has several important computational and conceptual advantages, as we will see. The free energy balance equation can be viewed as the instability process from the thermodynamic viewpoint. The generality of this relation means that it pertains to all the classes of instabilities usually considered, e.g., curvature driven modes, modes driven by parallel resonances and drift resonances, trapped particle modes, collisionally driven modes, etc, etc. One particular application of this equation is to allow us to consider how the charge flux constraint affects the instability process \cite{kotschA} without limiting our framework to a specific instability type. And, if our focus is on any specific instability (e.g. the ITG/TEM as in \cite{kotschA}), to consider that mode from the most general viewpoint: essential thermodynamic and statistical mechanical concepts. The effects of the charge flux constraint (FC) are often crucially important to weaken instabilities in transport barriers, where the gradients are strong but the transport is astonishingly weak. Transport barriers (TBs) are both an extraordinary phenomenon observed in existing devices, and, are important in future devices to obtain thermonuclear energy gain. 

One thermodynamic definition of free energy is the capacity to do work at a fixed temperature (Helmholtz free energy), given that, of course, entropy cannot decrease. In the gyrokinetic system, the presence of the gradients of the Maxwellian is the source of free energy that imbues it with such a capacity for work. A crucial manifestation of this capacity is the system performing work on fluctuations so that they grow, i.e, the presence of instabilities. The very strong gradients of TBs should enable a large amount of work to be performed on the fluctuations so that they grow rapidly to large amplitude. And for most physical systems, this is precisely what happens when there are steep gradients.  Instabilities growth is possible because, simultaneously, the fluctuations produce transport fluxes that acting to reduce the free energy in the gradients. The stronger the instability, the more rapid the relaxation. This is just an instance of the universal tendency of systems to minimize free energy, and is the most general perspective from which to examine instabilities. 

One of our goals is to understand how it is that extremely large gradients can exist in TBs, and yet, this large free energy is apparently unable to cause rapid instabilities to give rapid relaxation of that free energy. For this, we start by deriving a free energy equation for the gyrokinetic equation. Since this point of view is not usually taken in the community for gyrokinetic instability dynamics, we will be slightly pedagogical.

We begin with the nonlinear electrostatic gyrokinetic equation, which describes fluctuations of the non-adiabatic part of the distribution function around a local Maxwellian $F_{Ms}$, for each species s, with a temperature $T_s$. Specifically, it describes the distribution of the positions of the center of gyromotion:

\begin{eqnarray}
&&\frac{\partial h_s}{\partial t}+\vec{v_{tot}}\cdot \vec{\nabla} h_s + C(h_s) = \nonumber \\
&&\frac{q_s }{T_s}(\frac{\partial <\phi>}{\partial t}F_{Ms}+<\vec{v_{E \times B}}>\cdot\vec{\nabla}F_{Ms})
\frac{h_s}{F_M} \bm{C} ( h_s) \nonumber \\
\label{gyrokt}
\end{eqnarray}

where $<>$ denotes the gyro-average,  $\vec{\nabla} F_{Ms}$ is the local gradient of the background Maxwellian, and $\vec{v_{tot}}$ includes the parallel motion, $\vec{E} \times \vec{B}$ drifts, curvature drift and grad B drift. Within the nominal gyrokinetic ordering that applies to the large majority of calculations done in the field,  $F_{Ms}$ is regarded as a constant in space in eq\eqref{gyrokt}. Similarly, the local gradientis are also taken as constant in the perpendicular direction. (And note $F_{Ms}$ is constant on a flux surface, so $\vec{\nabla} F_{Ms}(\psi)=\partial F_{Ms} / \partial \psi\vec{\nabla}\psi $, where $\psi$ is a flux function) 

We multiply eq\eqref{gyrokt} by $h_s/f_{Ms}$ and integrate over $\vec{x}$ and $\vec{v}$. With conventional boundary conditions the $\vec{v_{tot}}\cdot \vec{\nabla}$ terms vanish. Furthermore, we define 

\begin{equation}
\delta f_s = -\frac{q_s <\phi>}{T_s} F_{Ms} +h_s
\label{distf}
\end{equation}

and also multiply by the species temperature $T_s$, and summing over species. We use the fact that $\int <a> b = \int a <b>$ (for conventional boundary conditions), to obtain the free energy balance equation

\begin{eqnarray}
&&\frac{\partial }{\partial t}  \left[ \sum_s  \frac{T_s}{2} \int \mathrm{d}\vec{x} \mathrm{d}\vec{v} \left( \frac {\delta f_s^2}{F_{Ms}} + (\phi^2-<\phi>^2)F_{Ms} \right)+\frac{|\vec{\nabla}_{\perp}\phi\ |^2 }{8\pi} \right] \nonumber \\
&&= \sum_s \big[\bm{n}_s (\bm{Q}_s \frac{1}{T_s} \frac{ d T_s}{d x} +
\Gamma_s \frac{T_s}{n_s} \frac{d n_s}{d x}) \big] - \sum_s \int d\bm{x} d\bm{v}
\frac{h_s}{F_M} \bm{C} ( h_s) \nonumber \\
&& + \int \mathrm{d}\vec{x} \phi \frac{\partial}{\partial t}(  \frac {\nabla_{\perp}^2 \phi}{4\pi}-\sum_s q_s \delta n_s ) \nonumber \\
\label{free1}
\end{eqnarray}

So far we have not imposed the condition that the fluctuations obey Maxwell's equations. But, the last term on the RHS vanishes when the fluctuations satisfy Poisson's equation, which in the gyrokinetic limit, is

\begin{equation}
\frac {\nabla_{\perp}^2 \phi}{4\pi}=\sum_s q_s \delta n_s
\label{poiseq}
\end{equation}

where the $\delta n_s$ are the perturbed densities, which are related to $h_s$ by

\begin{equation}
\delta n_s=\int \mathrm{d}v  <h_s> -  \frac{q_s \phi}{T_s}
\label{dispreln}
\end{equation}

For the typical ITG/TEM, Poison's equation becomes quasi-neutrality, since the LHS of eq\eqref{poiseq} is smaller than the RHS by $\sim k_{\perp}\lambda_{Debeye}^2 << 1$. However, for ETG modes, this may not be case. It will be instructive to not take the limit of small Debeye length, and hence, include the field energy in eq\eqref{free1}. Although last term in eq\eqref{free1} vanishes for fluctuations that satisfy Poisson's equation, it will prove useful to include it nonetheless, for some conceptual and computational reasons. 

This is a fundamental thermodynamic relation for the gyrokinetic system, so we discuss it's interpretation. 

In the limit $k_{\perp}^2 \to 0$, this approaches the eq(2) of \cite{kotschA} (for fluctuations that obey Poisson's equation). The apparently obscure term $\sim (\phi^2-<\phi>^2)$ in eq\eqref{free1} becomes the perpendicular kinetic energy from $E \times B$ motion in this limit, so it can be considered to be a part of the fluctuation energy. 

\begin{eqnarray}
&&\frac{\partial}{\partial t} \sum_s \big[ \frac{T_s}{2} \int d\bm{x} d\bm{v}
\frac{\delta f_s^2}{F_M} + \frac{\bm{m}_s \bm{n}_s \delta V^2_{E\times B}}{2}
+ \frac{\delta E^2}{8 \pi} \big] =  \nonumber \\
&&\sum_s \big[\bm{n}_s (\bm{Q}_s \frac{1}{T_s} \frac{ d T_s}{d x} +
\Gamma_s \frac{T_s}{n_s} \frac{d n_s}{d x}) \big] - \sum_s \int d\bm{x} d\bm{v}
\frac{\delta f_s}{F_M} \bm{C} (\delta f_s) \nonumber \\
\label{eq:one}&& 
\end{eqnarray}

As a pedagogical exercise, let us see that this equation can clearly be interpreted as a free energy balance for fluctuations about a Maxwellian plasma at a given temperature. The gyrokinetic equation applies to small fluctuations away from a Maxwellian with temperature T , $f=f_M+ \delta f$. The perturbed entropy $\delta S$ due to small fluctuations $\delta f$ is

\begin{equation}
\delta S = -\delta (f logf) = -\delta f \ log f_M - \frac {\delta f^2}{2 f_M} + H.O.T.
\end{equation}

(So at constant energy and particles, the maximum entropy state is a Maxwellian, $\delta f$=0, noting that $ log f_M=-m_s v^2/2T_s + Constant$). There is also a contribution to the total entropy production in the equilibrium, and the change in this entropy is the usual product of thermodynamic forces and thermodynamic fluxes on the RHS of  eq\eqref{eq:one}. From Boltzmann's H-theorem there is also entropy production from the collision operator $\bm{C}$.  Furthermore, the U in the Helmholtz free energy should include the field energy as well as the perturbed kinetic energy.  Finally, since the gyrokinetic equation conserves particles for each species, no changes in the free energy result from terms $\sim Constant\ \delta f$)

Taking all this into account, the LHS of eq\eqref{eq:one} corresponds to the rate of change of the free energy of fluctuations for all species summed. A crucial feature of eq\eqref{free1} is that the LHS is the derivative of a positive definite quantity. There is an entropy "cost" to increase the fluctuations away from a Maxwellian $\delta f$, and also an energy "cost" from the other terms on the LHS, and this is "paid for" by the decrease in the equilibrium free energy(relaxing the gradients). The eq\eqref{free1} has corrections due to gyro-averaging, since $h_s$ is for a distribution function of gyro-centers, but it is otherwise, clearly, essentially the same equation. 

One might also increase the fluctuations in the system, and hence its free energy, by doing work on the system by an external agent. The last term on the RHS of eq\eqref{free1} is potentially such a term. If Poisson's equation was not satisfied for the plasma charges, we could presume that there must be additional hypothetical external charges $\rho_{external}$ that bring Poisson's equation into balance $\frac {\nabla_{\perp}^2 \phi}{4\pi}-\sum_s q_s \delta n_s =\rho_{external}$. Using elementary electrodynamics, it is easily shown that the last term on the RHS is equal to the work done on the gyrokinetic charges by the external charges. So as expected, the free energy is increased by the amount of external work done on the system. This is the electrodynamic equivalent of a piston in elementary thermodynamics. 

For the usual plasma instabilities this external term vanishes. The system itself does the work on the fluctuations to cause them to grow. This comes at the expense of decreasing the free energy of the background, which is given by the usual terms with products of thermodynamic forces and corresponding fluxes. The equilibrium gradients act as a thermodynamic "potential energy" that is tapped into to increase the free energy in the fluctuations. 

The free energy balance equation will be satisfied as long as the last term in eq\eqref{free1} vanishes:

\begin{equation}
\int \mathrm{d}\vec{x} \phi \frac{\partial}{\partial t}(  \frac {\nabla_{\perp}^2 \phi}{4\pi}-\sum_s q_s \delta n_s ) =0.  
\label{realdr}
\end{equation}

For eigenmodes that satisfy Poisson's equation, this obviously holds. Equation\eqref{realdr} means that fluctuations do no net electrostatic work on themselves. Or, in other words, the fluctuation growth is sustained without any "assistance" from external agents doing additional work on the system. The growth rate is determined purely by how efficiency the fluctuations tap the thermodynamic potential energy (free energy) of the equilibrium gradients.  
 
Equation\eqref{realdr} is usually much easier to compute than the free energy balance equation eq\eqref{free1}. For fluctuations that satisfy the gyrokinetic equation, the vanishing of eq\eqref{realdr} is equivalent to the free energy equation being satisfied. 

This is a thermodynamic description of the instability growth process.

\section{The insufficiency of free energy balance for determining instabilities, and the need to consider the flux constraint}

What is the connection of the free energy equation to the usual eigenvalue problem? One considers linear gyrokinetic fluctuations that evolve in time as $\sim e^{-i\omega t}$. The usual approach is to solve the gyrokinetic equation for given $\omega$ and insert this into Maxwell's equations (here, Poisson's Equation), and solve for the value of $\omega$ that allows a nontrivial solution. 

Suppose we knew what the spatial structure of the eigenmode was, or, had a reasonable approximation to it. Using it, we could compute $\delta n_s$ as a function of the complex $\omega$ by using the gyrokinetic equation eq\eqref{gyrokt}. Then, an instructive integral of Poisson's equation, when written for the usual complex $\phi$ and $\delta n$, and the complex conjugate  $\phi^{*}$, is

\begin{equation}
Re[ \int \mathrm{d}\vec{x} \phi^{*} (  \frac {\nabla_{\perp}^2 \phi}{4\pi}-\sum_s q_s \delta n_s )] =0.  
\label{PoisRe}
\end{equation}

\begin{equation}
Im[ \int \mathrm{d}\vec{x} \phi^{*} (  \frac {\nabla_{\perp}^2 \phi}{4\pi}-\sum_s q_s \delta n_s )] =0.  
\label{PoisIm}
\end{equation}

Thus we have two real equations in two real unknowns ($\omega_r$ and $\gamma$), which constitutes a full dispersion relation to determine them.

Let us compare this to the free energy balance equation, which when written for the usual complex $\phi$ and $\delta n$, and the complex conjugate  $\phi^{*}$ , eq\eqref{realdr} is 

\begin{equation}
Re[\int \mathrm{d}\vec{x} \phi^{*} i \omega (  \frac {\nabla_{\perp}^2 \phi}{4\pi}-\sum_s q_s \delta n_s )] =0.  
\label{realdr2}
\end{equation}

\emph{But crucially, this is only one real equation, a linear combination of eq\eqref{PoisRe} and eq\eqref{PoisIm}. To determine the eigenfrequency, we need two relations, not just one }

\emph{ The crucial physical consequence is: considerations of free energy balance, although they are absolutely fundamental to the physics of the instability process, are not sufficient to determine the realizable eigenfrequencies.} 

The flux constraint can provide the other needed condition. In the case of an instability, using the quasilinear fluxes (averaged over long space scales), and with some manipulation, the FC is:

\begin{equation}
\sum_s q_s \Gamma_{rs} = 0 
\label{eqFC}
\end{equation}

After some manipulation, if can be shown that the charge flux constraint eq\eqref{eqFC} for exponential eigenmodes is equivalent to 

\begin{equation}
Im[ \int \mathrm{d}\vec{x} \phi (  \frac {\nabla_{\perp}^2 \phi}{4\pi}-\sum_s q_s \delta n_s )] =0.  
\label{Imdr}
\end{equation}

Together, eq\eqref{realdr} and eq\eqref{Imdr} are equivalent to eq\eqref{PoisRe} and eq\eqref{PoisIm}. (As long as the growth rate $\gamma$ does not vanish.) 

\emph{Thus, the free energy balance together with the flux constraint can constitute a dispersion relation.} 

As described in detail in \cite{kotschB} and  \cite{kotschC}, the physics of the FC is totally different from free energy dynamics, and can be interpreted as akin to the constraint of local momentum conservation of localized fluctuations. 

\emph{A crucial physical conclusion is that the FC can be a serious constraint upon the realizable free energy dynamics, as discussed in the section below.} Even if the free energy balance has solutions for growing modes, the FC might not have a solution for $\gamma>0$. 

As shown in \cite{kotschA}, there are circumstances where the FC is insoluble. That work is dedicated to explicating these concepts for the ITG/TEM modes, in detail. As described there, stronger density gradients tend to bring this situation about, because of the basic statistical mechanical fact that thermodynamic forces drive their respective thermodynamic fluxes. So stronger density gradients can drive the fluxes to be too strong to balance in the FC eq\eqref{eqFC}, if the dynamics of electrons and ions are sufficiently different.  

\emph{And such a circumstance has no physical connection to the amount of free energy in the equilibrium, i.e, how large $\gamma$ might be capable of reaching by only considering the free energy balance. }

It is straightforward to include electromagnetic effects in these considerations, and they leave the essential structure just described intact. Including the parallel vector potential $A_{\parallel}$, the result is 

\begin{eqnarray}
&&\frac{\partial }{\partial t} \Bigg[  \quad \sum_s  \frac{T_s}{2} \int \mathrm{d}\vec{x} \mathrm{d}\vec{v} \left( \frac {\delta f_s^2}{F_{Ms}} + (\phi^2 - \langle \phi \rangle^2)F_{Ms} \right) 
\nonumber \\
&&  \ \qquad \qquad \qquad \qquad \qquad +\quad  \Scale[1.4]{  \frac{|\vec{\nabla}_{\perp}\phi\ |^2 +B_{\perp}^2}{8\pi} } \qquad  \qquad \Bigg]  \nonumber \\
&&= \sum_s \big[\bm{n}_s (\bm{Q}_s \frac{1}{T_s} \frac{ d T_s}{d x} +
\Gamma_s \frac{T_s}{n_s} \frac{d n_s}{d x}) \big]      
- \sum_s \int d\bm{x} d\bm{v}\frac{h_s}{F_M} \bm{C} ( h_s) \nonumber \\
&& + \int \mathrm{d}\vec{x}  \phi \frac{\partial }{\partial t}(  \frac {\nabla_{\perp}^2 \phi}{4\pi}-\sum_s q_s \delta n_s ) \nonumber \\
&& + \int \mathrm{d}\vec{x}  \frac{A_{\parallel}}{c} \frac{\partial }{\partial t}(  \frac {\nabla_{\perp}^2 A_{\parallel}}{4\pi c}-\sum_s \delta j_{\parallel s} )
  \nonumber \\
\label{freeB}
\end{eqnarray}

Magnetic field energy is now included in the free energy of the fluctuations on the LHS. On the RHS, the fluxes now include electromagnetic contributions from the $A_{\parallel}$ terms, in addition to the electrostatic fluxes from ExB fluctuation. Such magnetic terms embody transport effects such as those from stochastic magnetic fields. And, if the fluctuations do not satisfy Ampere's Law, there is additional electromagnetic work done on the system by the last term of eq\eqref{freeB}.

For these magnetic contributions too, free energy consideration are only the real component of the last complex quantity in eq(\eqref{freeB}) which is obviously closely related to Ampere's Law, but are only a real part. In other words, once again, free energy considerations alone are insufficient to imply that Ampere's Law is satisfied. As in the in the electrostatic case, an additional relation is needed. This additional condition is provided by the charge flux constraint \emph{due to electromagnetic part of the fluxes alone. This can be shown to vanish, separately from the electrostatic component of the charge flux constraint} (see \cite{kotschB}). And just like the electrostatic FC, this necessary condition might also be a serious constraint upon the realizable free energy dynamics. (For example, as shown in \cite{kotschC}, for tearing mode like fluctuations, this magnetic part of the FC essentially requires that the real frequency be of order $\omega^{*}$.)

We note a final small detail in closing. One often considers gyrokinetic instabilities in the ballooning limit. To apply eq\eqref{free1} or eq\eqref{freeB} to such cases, note that this is a flux tube geometry, where modes only depend upon the length along a field line $l$. The volume element in coordinate space $\mathrm{d}\vec{x} \to \mathrm{d}l A \sim  \mathrm{d}l /B$ since the area perpendicular to a flux tune is $\sim 1/B$.

\section{Application of these considerations to a Simplified Kinetic Model}

Often, approximate 0D dispersion relations are considered for electrostatic modes.  An example is the Simplified Kinetic Model of \cite{kotschA}. Even though the geometry of SKiM is simplified, it is still a version of the gyrokinetic equation. Hence, one can show that the free energy balance equation is still obeyed, and, the flux constrain is also obeyed. Quasinuetrality gives the dispersion relation eq\eqref{dispreln}, and for such local dispersion relations, we denote this as $D(\omega)=0$. 

From eq\eqref{free1}, the vanishing of the LHS of the free energy balance equation is equivalent to  

\begin{equation}
Re(i\omega D(\omega))=0
\label{SKiMFEB}
\end{equation}

What about the FC? In the gyrokinetic equation, guiding centers move by the gyroaveraged ExB drift. This velocity, for the SKiM case, in the radial direction, is $i (c/B) k_y  J_0 (< k_{\perp} > \rho_i)\phi $. The total flux of gyrocenters is the product of this times the non-adiabatic part of the distribution function for each species $h_s$, integrated over velocity. The FC implies that the sum over all species of the charge flux from this vanishes, or

\begin{equation}
\sum_s Re[-i (c/B)q_s \int \mathrm{d} v \,  k_y  J_0 (< k_{\perp} > \rho_i)\phi^* h_s ]=0
\label{fluxgyro}
\end{equation}

Using typical 0D expressions for $h_s$ (as in \cite{kotschA}) this becomes 

\begin{equation}
Im(D(\omega))=0  
\label{SKiMFC}
\end{equation}

which is eq(\ref{Imdr})  for simplified models. As long as the growth rate does not vanish, these are two separate relations. When $\gamma > 0$, these dual equations, together, are equivalent to $D(\omega)=0$. As described above, the free energy balance together with the flux constraint specifies the mode frequency $\omega=(\omega_r,\gamma)$. 

The free energy balance describes how fast the fluctuation can potentially grow, given the free energy in the equilibrium. On could construct a graph, in the upper half plane of $\omega$ space, of solutions of the free energy balance. It would be a continuous path of points $(\omega_r,\gamma)$ with positive growth rates, if the system contains sufficient free energy so that growth is \emph{possible}. And quite likely, the peak growth rate for this curve could be very large when gradients are also strong. This is the situation in a TB, for example. 

Another curve in the upper half plane of  $\omega$ space gives the solution of the FC. The actual mode frequency is at the intersection of this curve and the free energy balance curve. 

There is no unstable mode unless \emph{unless the FC is also satisfied}. And, as described in \cite{kotschA}, there are generic circumstances when the FC becomes insoluble. So even though there is strong free energy, there will be no instability. Or, there are circumstances when the FC only has solutions for very low growth rates. Then, no matter where the intersection of the FC  and free energy balance occurs (if at all), the mode can, at best, only have a low growth rate.  

As shown in \cite{kotschA}, this is exactly the situation that allows stability of the ITG/TEM in a TB without velocity shear. There, one finds numerous graphs that show the situation described in the paragraphs above. 

These considerations are not limited to the ITG/TEM. The same basic structure applies to quite different modes, including strongly electromagnetic ones. The circumstances described above can be quite generic. If there are strong gradients in the equilibrium, the RHS of eq\eqref{freeB} is large. In other words, solutions of the free energy balance will likely trace a curve in the upper half of the $(\omega_r,\gamma)$ plane which reach large $\gamma$ for some $\omega_r$. But such growth rates may not be realizable, because the restriction of the FC always applies. It can still be quite possible that the FC has no solution, or, only has a solution at small $\gamma$. As described in \cite{kotschA}, such a circumstance tends to arise generically, when the density gradient is large, and when the physical dynamics of the ions and electrons are very different so that each tends to produce a very different charge flux. 

\bibliography{apssamp}
\bibliographystyle{apsrev4-1}

\end{document}